\documentclass[aps,prc,twocolumn,superscriptaddress,floatfix,longbibliography]{revtex4-1}%
\usepackage[utf8]{inputenc}
\usepackage{indentfirst}
\usepackage{tabularx} 
\usepackage{amsmath}
\usepackage{color}
\usepackage{ulem}
\usepackage[ bookmarks,
             bookmarksopen = true,
             bookmarksnumbered = true,
             linktocpage,
             colorlinks = true,
             linkcolor = blue,
             urlcolor  = blue,
             citecolor = magenta,
             anchorcolor = green,
             hyperindex = true,
             hyperfigures]{hyperref}

\usepackage{natbib}
\usepackage{graphicx}
\makeatother

\begin{document}

\title{Probing criticality with deep learning in relativistic heavy-ion collisions}

\author{Yige Huang}
\affiliation{Key Laboratory of Quark and Lepton Physics (MOE) \& Institute of Particle Physics,Central China Normal University, Wuhan 430079, China}
\author{Long-Gang Pang} 
\email{lgpang@ccnu.edu.cn}
\affiliation{Key Laboratory of Quark and Lepton Physics (MOE) \& Institute of Particle Physics,Central China Normal University, Wuhan 430079, China}
\author{Xiaofeng Luo}
\email{xfluo@ccnu.edu.cn}
\affiliation{Key Laboratory of Quark and Lepton Physics (MOE) \& Institute of Particle Physics,Central China Normal University, Wuhan 430079, China}
\author{Xin-Nian Wang}
\email{xnwang@lbl.gov}
\affiliation{Key Laboratory of Quark and Lepton Physics (MOE) \& Institute of Particle Physics,Central China Normal University, Wuhan 430079, China}

\affiliation{Nuclear Science Division, Lawrence Berkeley National Laboratory, Berkeley, CA 94720, USA}

\begin{abstract}
Systems with different interactions could develop the same critical behaviour  due to the underlying symmetry and universality.
Using this  principle of universality, we  can embed critical correlations modeled on the 3D Ising model into the  simulated data of heavy-ion collisions,  hiding
weak signals of a few inter-particle correlations within a large particle cloud.
Employing a point cloud network with dynamical edge convolution, we are able to
identify events with critical fluctuations through supervised learning,
and pick  out a large fraction of  signal particles used for decision-making in each single event.
\end{abstract}

\maketitle

\section{Introduction}
Quantum Chromodynamics (QCD) is the fundamental theory of the strong interaction. Exploring the phase structure of strongly interacting QCD matter is one of the main goals of heavy-ion collision experiment~\cite{Fukushima:2010bq,Bzdak:2019pkr,Luo:2017faz}. Lattice QCD~\cite{Aoki:2009sc,HotQCD:2019xnw,Ding:2015ona} predicts a smooth crossover transition from normal hadronic phase to Quark-Gluon Plasma (QGP) around temperature $T_c$=156 MeV at vanishing baryon chemical potential ($\mu_{B}$ = 0 MeV). At finite baryon density region, QCD-based models calculations~\cite{Shi:2014zpa,Gao:2016qkh,Fischer:2018sdj,Fu:2019hdw} indicate that there is a possible QCD critical point (CP), which is the end point of the first-order phase transition boundary between the hadronic matter and QGP. 

Searching for the CP is one of the most important goals in beam energy scan (BES) program at the Relativistic Heavy-ion Collider (RHIC) ~\cite{Fukushima:2010bq,Bzdak:2019pkr,Luo:2017faz}. Many theoretical and experimental efforts have been made to locate the CP~\cite{Stephanov:2004wx,Stephanov:2006zvm,Luo:2017faz}. One avenue is to classify  the smooth crossover and first order phase transition using  the information from the final state particle spectra and collective flow~\cite{Hofmann:1976dy,Stoecker:1986ci,Brachmann:1999xt,Brachmann:1999mp,Csernai:1999nf,Ivanov:2000dr,Rischke:1995pe,Stoecker:2004qu,Csernai:2004gk,Nara:2016hbg,Nara:2017qcg,Nara:2018ijw,Paech:2003fe}. This method  looks for the consequences of the softening of the equation of state  since the  pressure gradients are much smaller in a medium with a first order phase transition than a smooth crossover transition, which leads to slower fluid acceleration and smaller transverse momenta of final state particles.
Another avenue is to  search for the enhanced fluctuations  when the system goes through the critical point.
 These includes, for example, fluctuations of conserved charges \cite{Stephanov:2008qz,Stephanov:2011pb,STAR:2010mib,STAR:2013gus,STAR:2014uiw,STAR:2017tfy,STAR:2020tga,STAR:2021iop}, hydrodynamic fluctuations \cite{Nahrgang:2011mg,Herold:2013bi,Plumberg:2017tvu}, fluctuations caused by spinodal instabilities \cite{Li:2015pbv,Scavenius:2000bb,Palhares:2010be,Herold:2013qda,Li:2016uvu,Chomaz:2003dz,Randrup:2003mu,Sasaki:2007db,Steinheimer:2012gc,Steinheimer:2013gla,Steinheimer:2013xxa} and enhanced light nuclei yield ratio due to baryon density fluctuations~\cite{Sun:2018jhg,Yu:2018kvh,Sun:2020pjz,Zhao:2021dka}.

Many critical phenomena in systems with different interactions can develop the same critical behaviour with a universality that is dictated by the symmetry of the systems and can be described by same critical exponents \cite{Wilson:1973jj}.
Lee and  Yang  proved that  the Ising model in a magnetic field and a lattice gas are mathematically  equivalent \cite{Lee:1952ig}.
Employing this universality, one can  therefore map the QCD equation of state to that given by a 3-dimensional Ising model  with the same universality class \cite{Lee:1952ig,Stephanov:2004wx,Pradeep:2019ccv,Karthein:2021nxe,Teaney:2021dsl,Bluhm:2020mpc} to study the QCD phase diagram.
The divergence of the correlation length near the critical point will lead to the critical opalescence and scaling invariant, which means that the systems are self-similar when the resolution changes.
One thus expects that particles from the freeze-out hyper-surface close to the critical point 
have multi-particle fractal structure in the momentum space \cite{Bialas:1988wc,Satz:1989vj,Hwa:1989vn,Antoniou:2000ms,Wu:2019mqq}. 
Experimentally, intermittency analysis has been proposed to probe the self-similarity and density fluctuations in heavy-ion collisions.  Though a non-trivial intermittency  phenomenon is observed recently by the NA61/SHINE experiment at CERN SPS \cite{NA49:2012ebu,Davis:2019ala,Davis:2019vnk} in Ar+Sc collisions at 150 AGeV,  the magnitude of background fluctuations is big and the power law scaling is not fully established. No intermittency signal is observed in C+C, Pb+Pb and Be+Be collisions with similar collision energies.
Critical Monte Carlo simulations suggest a maximum critical proton fraction smaller than $0.3$\% in Be+Be collision,  indicating that traditional intermittency analysis may fail in looking for the weak signal of self-similarity, if the fraction of CMC particless is small compared with uncorrelated background .  It is interesting to explore whether the state-of-the-art deep learning can help to identify the weak intermittency signal from each event of heavy ion collisions.

Recently deep learning  has been used to study the QCD equation of states by classifying phase transition types,
using convolution neural network \cite{Pang:2016vdc,Pang:2021vwl,Du:2019civ,Kvasiuk:2020izb} and point cloud network \cite{Steinheimer:2019iso,Kuttan:2020mqj}. 
In heavy ion collisions  at low energies, auto-encoder with a single latent variable is  also used to study the order parameter of the nuclear liquid-gas phase transition \cite{Wang:2020tgb}.
In these studies, deep learning is powerful in mapping momentum or charge distributions of particles to the type of QCD phase transitions.
In this study, we will train a dynamical edge convolution network plus  a point cloud network to
identify weak intermittency signals of critical fluctuations, from exotic uncorrelated background particles. Employing Critical Monte Carlo (CMC) \cite{Antoniou:2000ms,Wu:2019mqq}, we encode the self-similarity in the inter-particle distances in momentum space. Further, we assume that only a small fraction of particles have intermittency which does not  change the single particle distribution.

This paper is organized as follows. In
Sec.II, we present the JAM transport model which is used to generate data  on multiple particle production in heavy ion collisions.
The CMC is used to generate intermittency signals of critical fluctuations
and the deep neural network is used for both classification and tagging.
In Sec. III, the prediction accuracy is compared for point cloud network and dynamical edge convolution neural network.
We also show the performance of signal-particle tagging.
In Sec. IV, we discuss and summarize the findings and the implications of the present  work.

\section{Method}
Probing critical fluctuations in heavy-ion collisions is a typical inverse problem. The information of criticality should be transmitted through the dynamical evolution of the dense medium in heavy-ion collisions and get encoded in the final state hadrons that are recorded by detectors. In the forward process, relativistic hydrodynamics as well as hadronic transport model are widely used to generate single particle distribution and multi-hadron correlations. In the present study, we use a hadronic transport model JAM~\cite{Nara:1999dz,Nara:2019crj} to generate background events without critical fluctuations. On the other hand, to introduce critical fluctuations, the so called Critical Monte-Carlo (CMC) model \cite{Antoniou:2000ms,Wu:2019mqq} is applied to generate a series of correlated particle momentum, which will be used to replace the momentum of particles in JAM events.

In the inverse process, a point cloud network and a dynamical edge convolution network are trained to identify critical fluctuations 
from large amount of uncorrelated background particles. The traditional intermittency analysis is also carried out to probe the encoded critical signals in the JAM events and validate the effectiveness of the deep learning method. 

\subsection{The JAM and Critical Monte-Carlo model}
JAM model is a hadronic transport model to simulate heavy-ion collisions~\cite{Sorge:1995dp,SORGE1997251,BASS1998255,Bleicher_1999,Kahana:1996ssp,LI1998556,Lin:2004en,Nara:1999dz,Nara:2019crj,Weil:2016zrk}. It simulates the complicated process from initial stage nuclear collisions to  multiple particle production and final state hadronic interactions. Independent binary collisions among hadrons including produced ones are modeled using the vacuum hadron-hadron scattering cross section.  In the present study, the mean field mode of JAM model is used to generate background events without including the critical fluctuations.

To simulate events involving critical fluctuations, Critical Monte-Carlo (CMC) model \cite{Antoniou:2000ms,Antoniou:2006zb,Wu:2019mqq} is used to generate a series of correlated particle momentum according to a power law function:
\begin{equation}
    f(\Delta p)=A \Delta p^{-\alpha}
\end{equation}
where $\Delta p$ is the distance of two CMC particles along an axis in momentum space.
$\nu=1/6$ is an index related to the universality class of Ising model, and we let $\alpha = 1 + \nu$.
$a$ and $b$ are the minimum and maximum of $\Delta p$, and in out study, we set $a=2\times 10^{-7} \mathrm{GeV/c}$ and $b = 2\mathrm{GeV/c}$.
$A=(\nu a^\nu b^\nu)/(b^\nu-a^\nu)$, is the normalization coefficient which is independent of $\Delta p$. In this study, we only consider 2D momentum space ($p_y,p_y$). 
The Levy flight random walk algorithm proposes the next step with strides respecting the distribution $f(\Delta p) = A\Delta p^{-\alpha}$ for $\Delta p_x$ and $\Delta p_y$ independently, and in this way, two sequence of $p_x$ and $p_y$ of CMC particles are generated whose adjacent differences $\Delta p$ obey the power law distribution. The self-similarity or intermittency is thus encoded in these CMC particles,
which is related to the observed large local density fluctuations associated with the critical point.

For such a probability density function $f(\Delta p)=A \Delta p^{-1-\nu}$ within a range of (a, b), it is possible to derive its cumulative distribution function: 

\begin{equation}
    F(\Delta p) = \frac{b^\nu(\Delta p^\nu-a^\nu)}{\Delta p^\nu(b^\nu-a^\nu)}
\label{cdf of p}
\end{equation}
where $F(\Delta p)$ is the cumulative distribution function of random variable $\Delta p$, $F(\Delta p)=\int_{a}^{b}{f(\Delta p)\mathrm{d}\Delta p}$. And one can then calculate the inverse function of $F(\Delta p)$: 

\begin{equation}
    \Delta p(F) = (\frac{a^\nu b^\nu}{b^\nu-b^\nu F + a^\nu F})^{1/\nu}
\label{inverse func}
\end{equation}
By randomly picking up a $F$ respecting to uniform distribution between 0 and 1, and using Eq. ~\ref{inverse func}, one can obtain a $\Delta p$.

\subsection{Data set preparation}
We generate about $2.2\times 10^5$ events of Au+Au central collisions at $\sqrt{s_\mathrm{NN}}$ = 27 GeV with impact parameters $b<3\ \mathrm{fm}$. Each event consists of hundreds of charged particles including pion, kaon and proton. The transverse momentum $p_x$ and $p_y$ are considered as two features of each particle. Therefore, each event stores one particle cloud in 2-dimensional momentum space. $2\times 10^5$ events are used to form the training set, while the number of events for validation and test are $1\times 10^3$ and $2\times 10^4$, respectively. 
For each JAM event, a corresponding CP event is created that encodes the critical fluctuation signals from CMC model. As a result, $4.4\times 10^5$ events in total are used in our study. To avoid data pollution,  event  with critical fluctuations and its corresponding JAM event are always put in the same data category. In this case, if one JAM event is in the training data, the event  with critical fluctuations associated with that JAM event is also put in the training data.  We will refer to these events with critical fluctuations as CP events and these particles encoded with the critical fluctuations as CMC particles.
Since the CMC model only generates the momentum correlation pattern and does not include the information of specific particle species, we don't distinguish between the types of particles when performing the replacement of particle in a JAM event.

For a given JAM event, we use replacing rate $\eta = N_{CMC}/N_{JAM}$ to describe the multiplicity ratio of CMC events to JAM events, the number of CMC particles introduced into its corresponding CP event can reflect how strongly the critical signal is encoded. In our study, two kinds of CP events with $\eta=5\%$ and $\eta=10\%$, respectively, are prepared. The detailed replacing procedures are listed below:

\begin{enumerate}
    \item Randomly select a particle in the chosen JAM event, use its $(p_x,p_y)$ as the starting momentum for generating the CMC event.
    \item Fill a histogram $H$ of the transverse momentum distribution from the generated CMC event. Remark the maximum magnitude of this histogram as $M$.
    
    \item Loop over the particles in the JAM events. For each particle, find its corresponding $p_{T}$ bin in $H$, record the content of $H$ in the $p_{T}$ bin as $f$.
    
    \item Get a random number $y$ in range from $0$ to $M$ respecting to uniform distribution. If $y \le f$, randomly select a CMC particle in the $p_T$ bin and replace this JAM particle with it; and if $y > f$, give up this JAM particle and go back to step 3 to find next JAM particle.
    
    \item Repeat step 3 to 4 until all the CMC particles are used or all the JAM particles are looped.
\end{enumerate}

By applying such algorithm, it is possible to keep the $p_T$ spectra of the substituted JAM particles close to that of the introduced CMC particles, hence the $p_T$ spectra of the JAM event and the corresponding CP event are quit similar. Even if there has a fluctuation of $p_T$ distribution, the overall $p_T$ spectrum will not be greatly affected due to the small fraction of CMC particles (5\% or 10\%) in the CP event. Considering the momentum resolution of experimental detector, we introduced a uncertainty for momentum of each particle in JAM event with a smearing as $\delta p_i \approx \pm 0.05p_i$, where $i=x,y$. The smearing operation will be done after the JAM and CP events are generated. 

\subsection{Intermittency analysis}
Local density fluctuations near the QCD critical point can be probed by intermittency analysis of scaled factorial moments~\cite{Wu:2019mqq} in relativistic heavy-ion collisions.
The scaled factorial moments (SFM)\cite{Wu:2019mqq} are defined as follows,
\begin{equation}
F_q(M)=\frac{\langle\frac{1}{M^D}\sum^{M^D}_{i=1}{n_i(n_i-1)\cdot\cdot\cdot(n_i-q+1)}\rangle}{\langle\frac{1}{M^D}\sum^{M^D}_{i=1}n_i\rangle^q}
\end{equation}
where $M$ is the number of grids in momentum space with equal size, $D$ is the dimension,
$i$ is the number of particles in the $i$th momentum-grid,
and $q$ is the order of the SFM method.

When $M$ is large, the power law dependence of SFM on the number of partitioned bins implies a self-similar correlations in the studied system\cite{Bialas:1985jb, Bialas:1988wc}.

\begin{equation}
    F_q(M)\approx (M^D)^{\phi_q}
\end{equation}

The intermittency index $\phi_q$ can characterize the strength of intermittency behavior and is related to the anomalous fractal dimension of the system\cite{DeWolf:1995nyp}. And there are studies show that using intermittency measurement together with the estimated freeze-out parameters can estimate the possible critical region of the QCD CEP\cite{Antoniou:2018kju}.

\begin{figure}[!htp]
\centering
\includegraphics[width=0.45\textwidth]{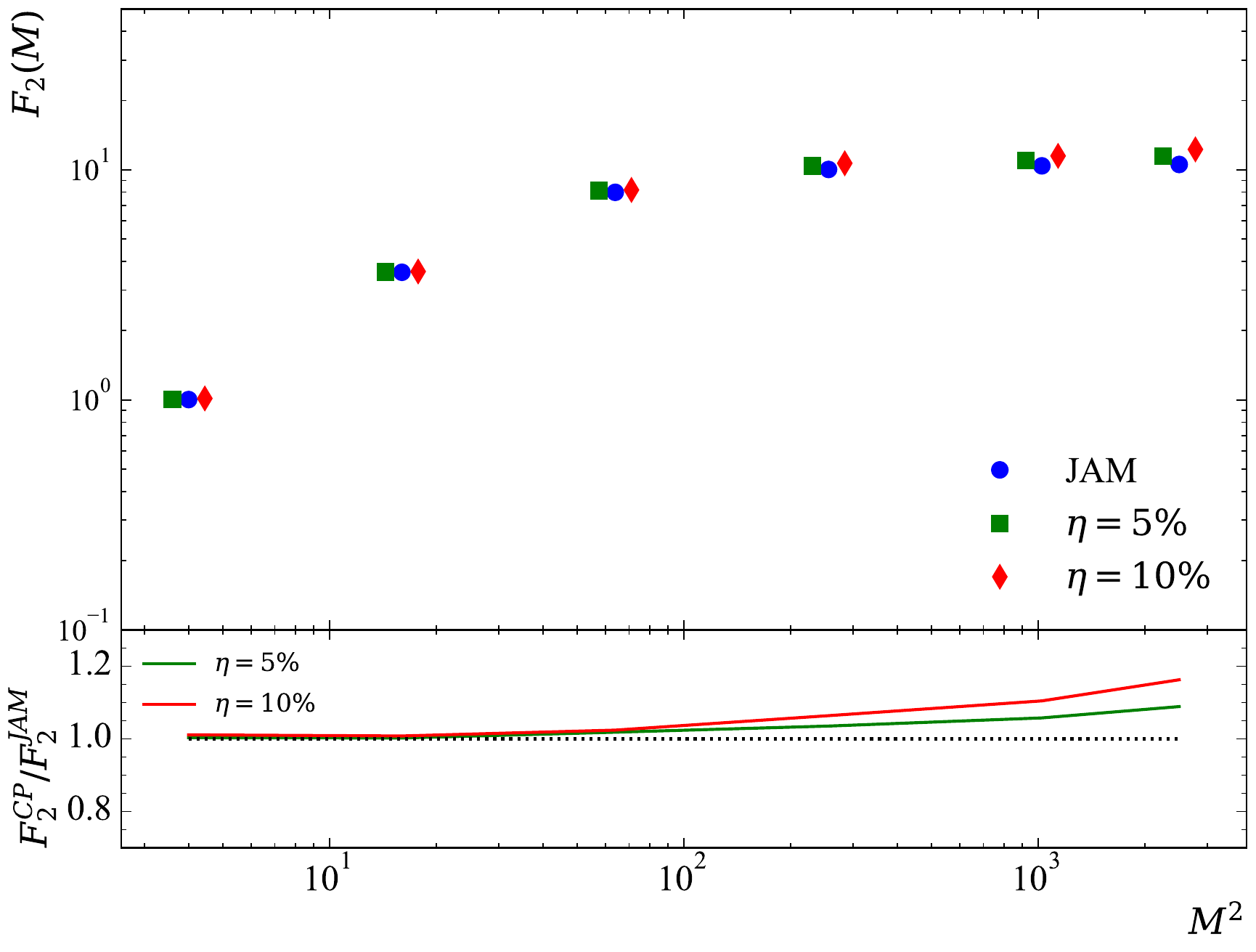}
\caption{The second order scaled factorial moments analysis for uncorrelated JAM events and events with critical fluctuations.
The upper-panel shows the absolute values of SFM for JAM events and events with 5\% and 10\% CMC particles.
To avoid the overlap of markers, results of critical events are slightly shifted horizontally for a clearer visualization. The lower-panel shows the ratios between critical and normal JAM events. 
No significant differences are observed for the absolute SFM values and their ratios.}
\label{fig: SFM}
\end{figure}
In the present study, the second order SFM ($q=2$) in two dimensional space ($D=2$) are studied 
for $M=$ 2, 4, 8, 16, 32, 50. 
As we take the experimental detectors into consideration, in SFM calculation, we only take no more than 50 grids for each dimension in a range of plus-minus 2.5 $\mathrm{GeV/c}$ to keep $p_T$ resolution to be like experimental options and at about 0.1 GeV/c.

As shown in Figure.~\ref{fig: SFM}, the intermittency analysis using the  SFM method \cite{NA49:2012ebu,Davis:2019ala,Davis:2019vnk,Wu:2019mqq} 
can not  differentiate  CP events  with 5\% and 10\% CMC particles that carry critical fluctuations from uncorrelated JAM events.

\subsection{Dynamical edge convolution neural network}

\begin{figure*}[!htp]
\centering
\includegraphics[width=1.0\textwidth]{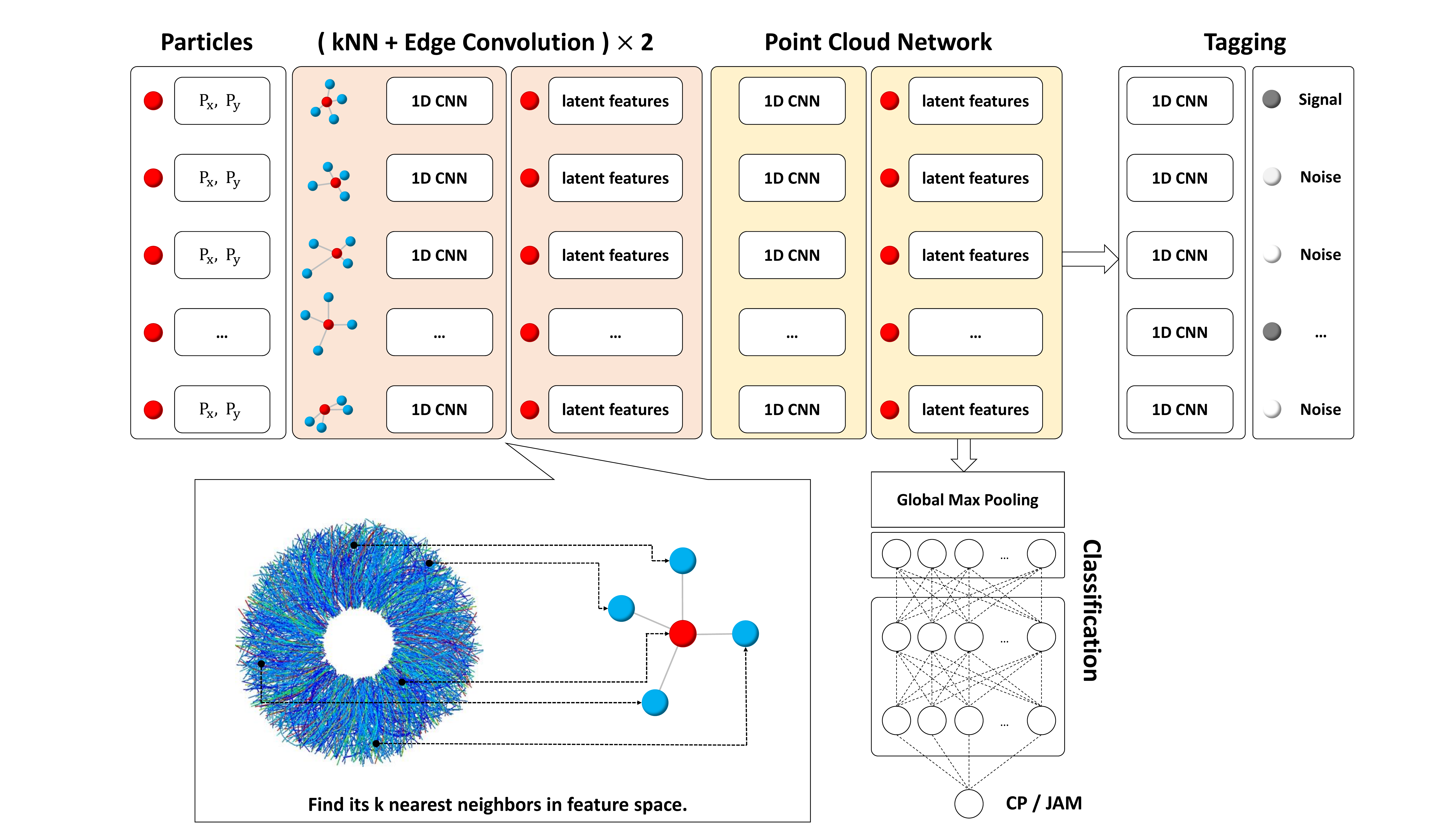}
\caption{Dynamical edge convolution neural network with point cloud module for both classification and tagging. The edge convolution block looks for k nearest neighbors of each particle to obtain a latent representation of that very particle, with short or long range correlations encoded deeply in. The representation of each particle are used in two tasks. One is the classification task to identify critical fluctuations from uncorrelated background events. The other is the tagging task to label correlated particles used for decision making.}
\label{fig:network}
\end{figure*}

A graph-based dynamical edge convolution neural network is trained for our multi-task learning. The input to the neural network are the particle cloud of each event, which consists of a list of particles with their information on $(p_x, p_y)$.
The output of the neural network corresponds to two tasks.
The first task is  the binary classification which requires true labels of each single event for supervised learning,
with CP indicating events with critical fluctuations and JAM indicating events without.
The second task is  the particle tagging which requires true labels of each single particle,
with 0 or 1 to indicate whether the particle is generated using Critical Monte Carlo model.

Shown in Figure.~\ref{fig:network} is the architecture of our neural network.
There are two kNN plus dynamical edge convolution blocks connecting to the input layer.
In the first block, kNN is used to find the k-nearest neighbors of each particle in $(p_x, p_y)$ space.
A fully connected network is used to learn edge features $\phi(\vec{p}_i, \vec{p}_{j})$ between the $i$'th
particle and its $j$'th neighbor. 
This module is shared by all its neighbors of particle $i$ to produced edge features
and that explains the name ''edge convolution''. 
The information of particle $i$ together with its edge features are feed to the second block.
Edge convolution layer would not only make use of the features of input neuron itself, but also take the relevance between the clustered units near that neuron into consideration, thus it can effectively capture the correlation information between particles.

The second kNN find the k-nearest neighbors of each particle in feature space.
It is thus possible to correlate particles that are far away in momentum space.
The neighbors of each particle change dynamically when the distances are computed in 
feature space, that is why the method is called ''dynamical edge convolution''.

The features of each particle together with its ''local'' information are flattened and
feed to a fully connected neural network to get a high dimensional latent variable for each particle.
The latent variable provides a high dimensional representation of each particle.
The above neural network is also shared by all particles and is called 1D convolution neural network (CNN).
Finally, the latent variables of each particle are used for two different tasks.
The module of ''Classification'' task is shown in the lower right corner.
A global max pooling gets the maximum values of each feature among all particles.
This  symmetric permutation operation learns the global feature of each particle cloud and
is used to determine whether it is a CP or JAM event.
The module of ''Tagging'' task is shown on the right of Figure.~\ref{fig:network}.
A 1D CNN with one output neuron is used to tag each particle in the particle cloud.
This module provides interpretation on whether the correlated particles are used
to identify events with critical fluctuations.
We have labeled correlated CMC particles as ''signal''
and uncorrelated JAM particles as ''noise''.
Binary cross entropy is used to compute the differences between the tagging output and 
the true labels of each particle.
The loss values of tagging module is added to the total loss with a weighting factor $10^{-3}$
such that the network focus more on ''classification'' task.

For comparison, we also train a point-cloud network without the kNN and dynamical edge convolution blocks shown in Figure.~\ref{fig:network}. The $(p_x, p_y)$ of each particle is directly feed to 1D CNN with
256, 128 and 64 channels respectively for classification.
Global average pooling layer is used in this simple point-cloud network as it performs better here.
Without kNN and dynamical edge convolution, the network can not capture much local information for intermittency identification. 

\section{Results and discussion}

\subsection{Classification accuracy}

\begin{figure}[!hptb]
\centering
\includegraphics[width=0.4\textwidth]{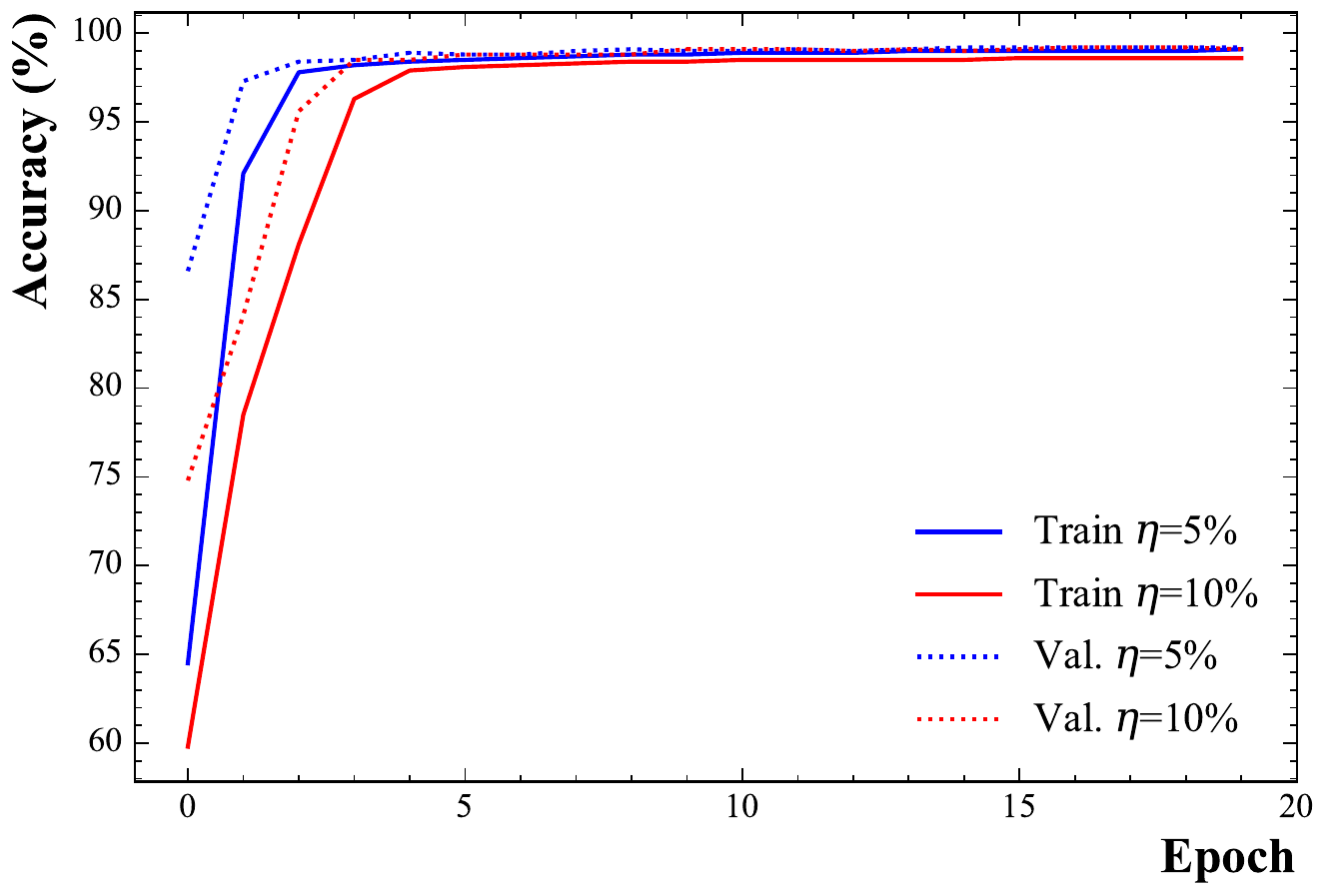}
\caption{The training and validation accuracy as a function of epochs. The training accuracy is in solid lines, for replacing rate $5\%$ (blue) and $10\%$(red). The validation accuracy is in dashed lines for replacing rate $5\%$ and $10\%$.}
\label{fig:train_valid}
\end{figure}

Shown in the Figure.~\ref{fig:train_valid} are the training (solid lines) and validation (dashed lines) accuracy as a function of training epochs.
Both training and validation accuracy increase as the model is trained longer with more epochs.
The validation accuracy reaches a maximum of 99.3\%, which means that deep learning is able to classify each single event with high accuracy, for
uncorrelated JAM events and events mixed with 90\% uncorrelated JAM particles and 10\% CMC particles ($\eta=10\%$).
For a smaller replacing rate ($\eta=5\%$), both validation and training accuracy decrease as compared with ($\eta=10\%$),
whose maximum value is about $93.3\%$.
Note: the smeared 5\% and 10\% both got 93.3\% acc. for validation set, while the 10\% one got higher score for test set.
The validation accuracy is slightly higher than training accuracy caused by the dropout and batch normalization layers
used in the network.
These two kinds of layers are known to be able to increase the generalization of the network by introducing noise during training.

Shown in Table.~\ref{tab:acc} are the testing accuracy of four different configurations.
Using the dynamical edge convolution plus point cloud network  we constructed in this study,
the testing accuracy are $97.7\%$ for $10\%$ replacing rate  and $92.8\%$ for $5\%$ replacing rate,
which are not quite far away from the validation accuracy.
Removing the dynamical edge convolution block, we have tested the performance of the point cloud network
with varying numbers of layers and neurons per layer to get the best testing accuracy.
The testing accuracy decreases to $84.8\%$ for $10\%$ replacing rate  and $ 83.4\%$ for $5\%$ replacing rate.

Another test set is prepared to make sure that the network make their decision based on multi-particle correlation in the CMC particles. In this test set, 5\% or 10\% particles of a JAM event are replaced by same amount of particles sampled randomly from many other events, one particle from each event to eliminate the two particle correlation in the replaced particles. If our network trained to identify CMC particles is fooled to classify these mixed events as CMC events, it means that the network learns the missing correlation in the replaced particles as compared with original JAM particles. In practice, our trained network treat these mixed events as JAM events, which is a proof that the network make their predictions using signals of CMC particles.

\begin{table}
    Testing accuracy
    \setlength{\tabcolsep}{6mm}{
    \begin{tabular}{ccc}
    \toprule
        $\eta$ & Edge-Conv  & Point-Cloud Net\\ \colrule
        $5\%$  & 92.8\% & 83.4\% \\ 
        $10\%$  & 97.7\% & 84.8\% \\ 
        \botrule
    \end{tabular}}
    \caption{The testing accuracy for dynamical edge convolution network and a simple point cloud network.}
    \label{tab:acc}
\end{table}

\subsection{Interpretability: tagging}

To figure out how does the network make its decision in identifying critical fluctuations from  the background,
we have added a tagging layer to the neural network.
To quantify the tagging performance, we introduce two metrics as follows,

\begin{equation}
r_{\rm c} = \frac{N_{C}}{N_{C}+N_{M}},\quad\;
r_{\rm t} = \frac{N_{C}}{N_{C}+N_{W}} 
\end{equation}
where $r_{\rm c}$ is the catching rate defined as the ratio between the number of correctly tagged particles $N_C$
and total number of signal particles $N_{C}+N_{M}$, 
where $N_{M}$ is the number of signal particles missed by the tagging module.
$r_{\rm t}$ is the tagging rate defined as the ratio between the number of correctly tagged particles $N_C$
and the total number of tagged particles $ N_{C}+N_{W}$,
where $N_W$ is the number of wrongly tagged uncorrelated particles.

The average catching rates $r_{\rm c} = 73.6\%$ for $\eta=5$\% and  $r_{\rm c} = 75.9$\% for $\eta=10$\%
indicate that the network may use about $3/4$ of the correlated particles to make its decision.
On the other hand, the tagging rate $r_t=94.5$\% for $\eta=5$\%  and  $r_t=95.4$\% for $\eta=10$\%
are much higher than catching rate $r_c$.
This result tells us that the tagging module can label CMC particles quite precisely. 

Since both edge convolution and the following 1D convolution layers of tagging module perform the same transformation for each particle, we can reversely track the tensor of labeled particles in the hidden feature space in the forward propagation process of neural network. 
For each input CP event, by checking the feature space after passing edge convolution layer, for a total of $N$ CMC particles well tagged, we find the $k$ nearest particles in the feature space corresponding to the feature vector of each particle, and count the number $M$ of CMC particles that were also well tagged. The proportion of those well tagged CMC particles from kNN to the total number of these kNN particles can then be calculated as $\frac{M}{k\times N}=94\%$. 
This result indicates that, the feature space transformation guided by edge convolution can aggregate CMC particles into a cluster in the new feature space, and then the tagging module can label them through the subsequent 1D convolution layers.

\begin{figure}[!hpbt]
\centering
\includegraphics[width=0.45\textwidth]{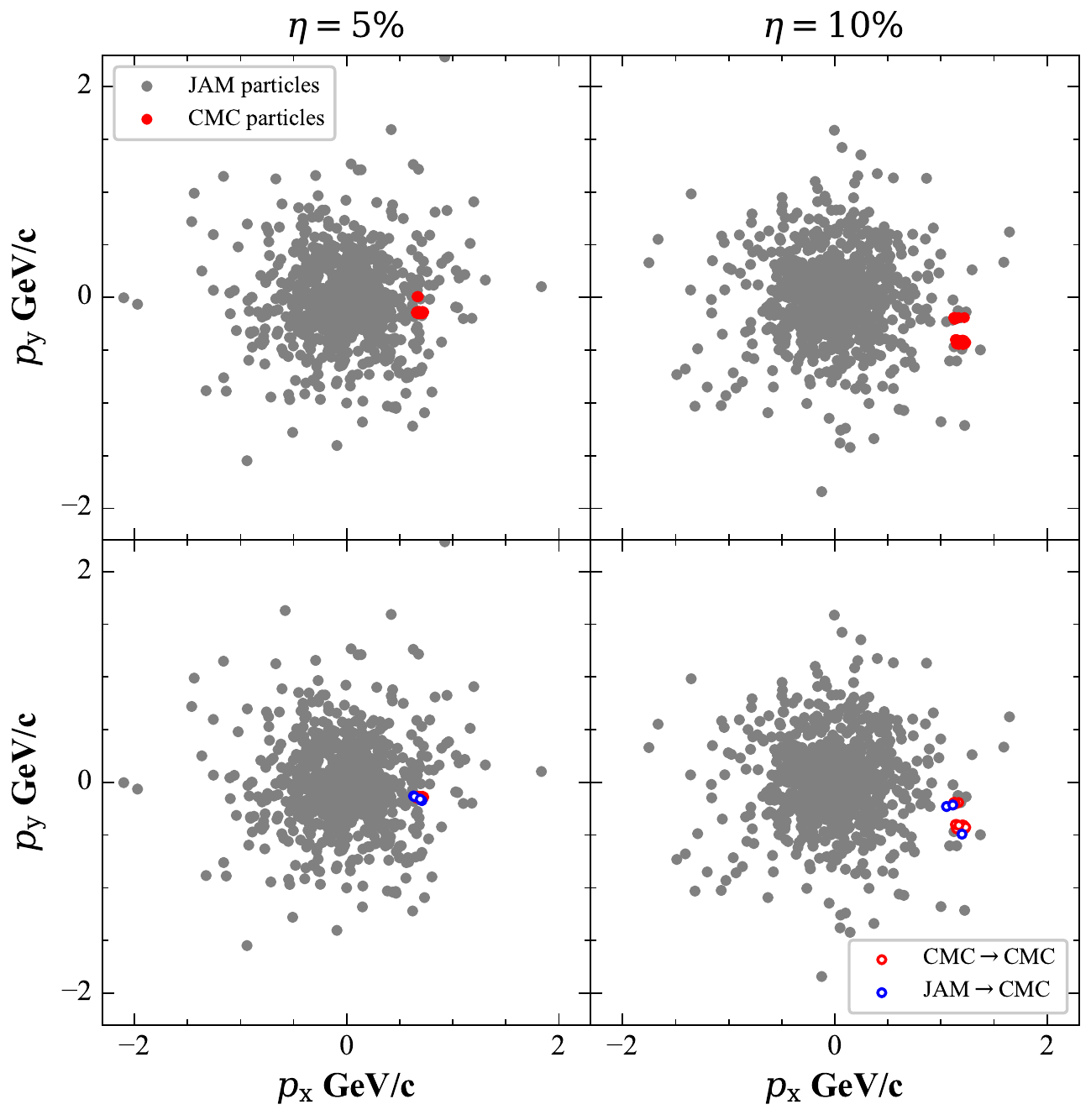}
\caption{The upper subplots show the comparison of JAM event and its corresponding CP event, in which the grey dots are the unchanged JAM particles, and the red ones are the critical  particles introduced by CMC events. The lower subplots are labeled results of tagging network, and the red dots refer to particles which were tagged correctly, while the blue ones are JAM particles labeled as CMC ones, while the grey dots are unlabeled particles. The graphs on the left show an example of $\eta=5$\%, while the ones on the right show an example of $\eta=10$\%. Although the CMC clusters in the two examples shown are all distributed on the right side of phase space, the location of CMC particles are not restricted indeed and they can be on any corner of the plot.}
\label{fig:tagging_demo}
\end{figure}

Figure~\ref{fig:tagging_demo} demonstrates the output of the tagging module.
In the upper subplots, grey dots represent unchanged JAM particles and red dots represent all the CMC particles in two testing events.
The corresponding tagging output for these two events are shown in the two lower subplots,
where the red dots represent CMC particles correctly tagged by the network while the 
blue ones are JAM particles but incorrectly tagged as CMC particles.
In average, $3/4$ of CMC particles are recognized by the tagging module.
And as discussed before, the incorrectly tagged particles are much fewer than correctly tagged CMC particles.
The two figures in the left are for $5\%$ replacing rate while the ones on the right are for $10\%$ replacing rate.

\begin{figure}[!hpbt]
    \centering
    \includegraphics[width=0.45\textwidth]{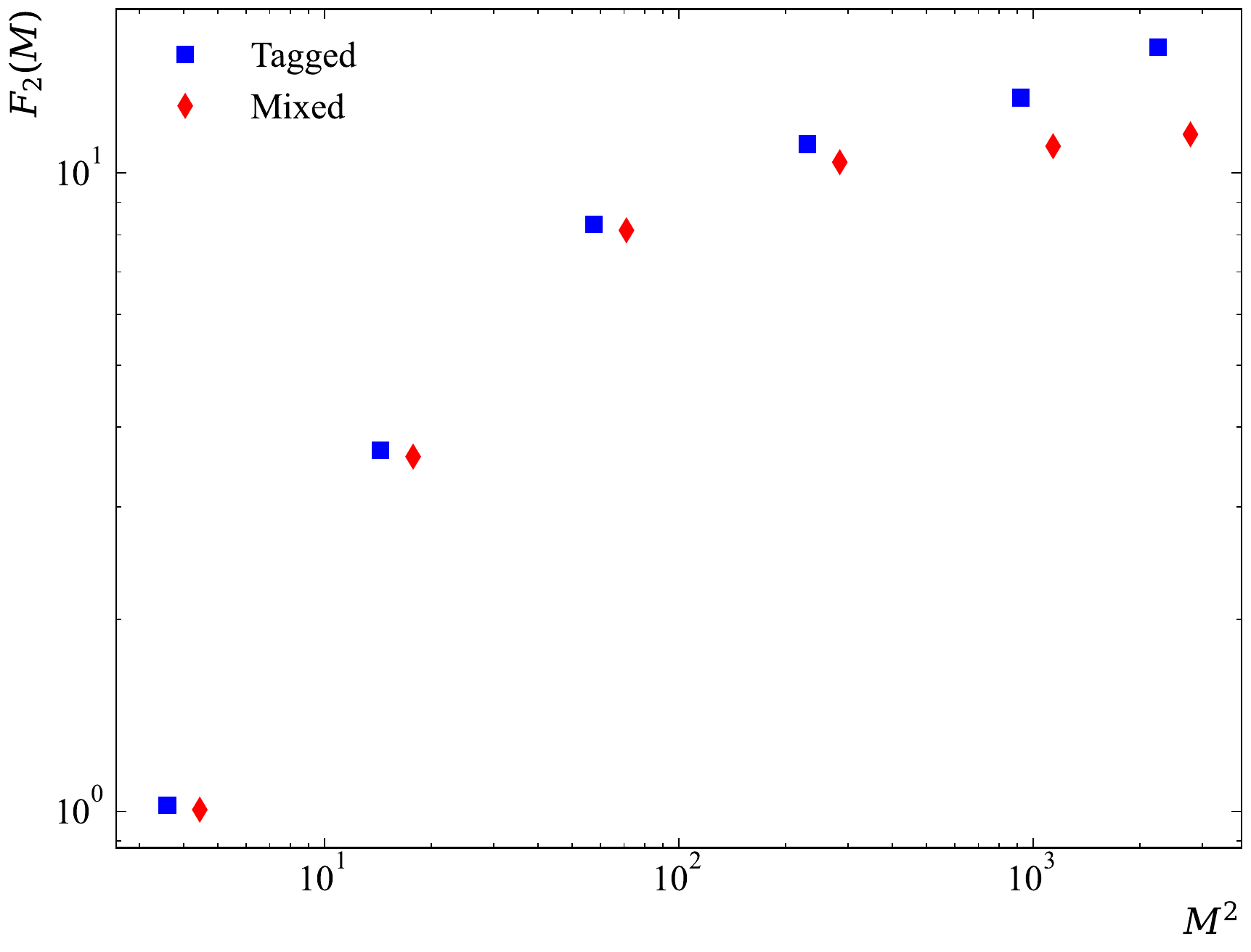}
    \caption{The 'Mixed' labeled red diamond markers represent the SFM results of all particles from $\eta$=5\% CP events, while the 'Tagged' labeled blue square markers stand for the SFM of tagged part of those events. As $M^2$ increase, the red diamonds have a flat performance, and the blue squares show a increment.}
    \label{fig: tagging sfm}
\end{figure}

Figure~\ref{fig: tagging sfm} shows the SFM calculation of $\eta=5\%$ CP events and the SFM of tagged particles of them, the former ones event have no increment with the increase of $M^2$ while the tagged ones present slight power law. This result reflects that the tagging module can somehow extract the encoded intermittency information.

\section{Summary and outlook}

In summary,  we have constructed a dynamical edge convolution plus point cloud network to identify the weak intermittency signal from the experimental data of heavy-ion collisions. We have demonstrated that such a state-of-the-art deep learning network enables us   to achieve a testing accuracy 92.8\% if only 5\% of JAM particles in each event are replaced
by correlated CMC particles.
The performance increases to $97.7$\% if the replacing rate of correlated particles increases to 10\%.
Removing the dynamical edge convolution block will decrease the performance by a large margin. Using tagging module, we further demonstrate that the network  can use around $3/4$ of correlated particles to make their decision.
At the same time, only about 5\% of uncorrelated background particles are incorrectly tagged as CMC particles.

We observe that the network can identify self-similarity or scaling invariant from   uncorrelated background.
This is important for experimental data analysis since only one indication of intermittency is observed in Ar + Sc collisions
whereas several other systems with similar collision energies fail.
Different from previous theoretical studies, we preserve the single particle distribution while introducing  a small fraction of particles with multi particle fractal structure.
This is more realistic but also   difficult   for the traditional intermittency analysis.
Based on our study, deep learning shows strong pattern recognition ability in identifying weak intermittency signals associated with critical phenomena. 
The method developed in this study can be applied to probe the critical fluctuations in heavy-ion collisions and can also be used to explore the criticality of other systems.

\section*{Acknowledgement}
We thank Jin Wu for helpful discussions on the critical monte carlo model.
 This work is supported by the National Key Research and Development Program of China (Grant No. 2020YFE0202002 and 2018YFE0205201), the National Natural Science Foundation of China under Grant Nos. 12122505, 11935007, 11221504, 11890711, 11861131009 and 12075098, and by the Director, Office of Energy Research, Office of High Energy and Nuclear Physics, Division of Nuclear Physics, of the U.S. Department of Energy (DOE) under grant No. DE- AC02-05CH11231, by the U.S. National Science Foundation under No. OAC- 2004571 within the X-SCAPE Collaboration. Computations are performed at Nuclear Science Computer Center at CCNU (NSC3). LG Pang and YG Huang also acknowledge the support provided by Huawei Technologies Co., Ltd.

\bibliography{references}

\end{document}